\documentclass[fleqn,useAMS,usenatbib]{mnras}
\usepackage{psfrag,color,epsfig}

\usepackage{graphicx}	
\usepackage{amsmath}	
\usepackage{amssymb}	
\usepackage{multicol}	
\usepackage{bm}		
\usepackage{pdflscape}	
\usepackage{multirow}
\usepackage{tabularx}  
\usepackage{caption}
\usepackage{newtxtext,newtxmath}

\usepackage[T1]{fontenc}
\usepackage{ae,aecompl}

\def\HI{\ion{H}{I}~}

\def\x{{\bm{x}}}

\usepackage{tikz,xcolor,hyperref}
\definecolor{lime}{HTML}{A6CE39}
\DeclareRobustCommand{\orcidicon}{%
	\begin{tikzpicture}
	\draw[lime, fill=lime] (0,0) 
	circle [radius=0.16] 
	node[white] {{\fontfamily{qag}\selectfont \tiny ID}};
	\draw[white, fill=white] (-0.0625,0.095) 
	circle [radius=0.007];
	\end{tikzpicture}
	\hspace{-2mm}
}
\foreach \x in {A, ..., Z}{%
	\expandafter\xdef\csname orcid\x\endcsname{\noexpand\href{https://orcid.org/\csname orcidauthor\x\endcsname}{\noexpand\orcidicon}}
}

\title[Exotic models with the dark ages 21-cm signal]{Constraining exotic dark matter models with the dark ages 21-cm signal}

\author[R. Mondal, R. Barkana \& A. Fialkov]{Rajesh Mondal\orcidA,$^{1,2}$\thanks{E-mail: rajeshmondal18@gmail.com}  Rennan Barkana\orcidB,$^2$ Anastasia Fialkov$^{3,4}$\\
$^{1}$Department of Physics, National Institute of Technology Calicut, Calicut 673601, Kerala, India \\
$^{2}$Department of Astrophysics, School of Physics and Astronomy, Tel Aviv University, Tel Aviv 69978, Israel  \\
$^{3}$Institute of Astronomy, University of Cambridge, Madingley Road, Cambridge, CB3 0HA, UK \\
$^{4}$Kavli Institute for Cosmology, Madingley Road, Cambridge, CB3 0HA, UK
}

\date{Accepted 2023 October 24. Received 2023 October 24; in original form 2023 September 05}
\pubyear{2023}

\begin{document}
\label{firstpage}
\pagerange{\pageref{firstpage}--\pageref{lastpage}}
\maketitle


\begin{abstract}
The dark ages 21-cm signal is a powerful tool for precision cosmology and probing new physics. We study two non-standard models: an excess radio background (ERB) model (possibly generated by dark matter decay) and the millicharged dark matter (mDM) model. These models were inspired by the possible EDGES detection of a strong global 21-cm absorption during cosmic dawn, but more generally they provide a way to anticipate the potential discovery space. During the dark ages the 21-cm global signal in the ERB model reaches a saturated form for an amplitude $A_{\rm r}=0.4$, where $A_{\rm r}$ is the radio background intensity at cosmic dawn relative to the cosmic microwave background. This amplitude is one fifth of the minimum required to explain the EDGES signal, and corresponds to just 0.1\% of the observed extragalactic background; it would give a signal that can be detected at 5.9$\sigma$ significance (compared to 4.1$\sigma$ for the standard signal) and can be distinguished from the standard (no ERB) signal at 8.5$\sigma$, all with a 1,000 hour global signal measurement. The 21-cm power spectrum has potentially more information, but far greater resources would be required for comparable constraints. For the mDM model, over a range of viable parameters, the global signal detection significance would be $4.7-7.2\,\sigma$, and it could be distinguished from standard at $2.2-9.3\,\sigma$. With an array of global signal antennas achieving an effective 100,000 hr integration, the significance would be 10$\times$ better. Our analysis helps motivate the development of lunar and space-based dark ages experiments.
\end{abstract}



\section{Introduction}
\label{sec:intro}
The 21-cm signal from neutral hydrogen (\ion{H}{I}) is the most promising method for studying a significant time period in the early Universe, from shortly after recombination through cosmic dawn and reionization. \citep{Sunyaev1972, Hogan, Scott1990}. In particular, an era that remains observationally unexplored is the dark ages, the period of time between the epoch of recombination (redshift $z\sim1100$) and the formation of the first luminous objects ($z\sim 30$). After recombination, the temperature of the cosmic microwave background (CMB)\,($T_{\gamma}$) fell as $(1+z)$, while the (kinetic) temperature of the gas\,($T_{\rm K}$) declined faster, eventually adiabatically as $(1+z)^2$. The spin temperature\,($T_{\rm S}$) was strongly coupled to $T_{\rm K}$ through collisional coupling until $z\sim70$ \citep{madau97}. After this time, the collisional coupling of the 21-cm transition became less effective, and $T_{\rm S}$ began to approach $T_{\gamma}$. Therefore, during the dark ages, the 21-cm signal from \HI is expected to be observable in the CMB spectrum over a wide redshift range of $200 \lesssim z \lesssim 30$. This is because $T_{\rm S}$ was significantly lower than $T_{\rm \gamma}$ during this time, which caused the \HI to absorb CMB photons.

The dark ages are a critical window into the history of the Universe. Unlike the present Universe, which is characterized by complex astrophysical processes, the dark ages offer a probe of fundamental cosmology. This is because the dark ages were a time when the Universe was still largely homogeneous and isotropic, and the only (currently known) source of radiation was the CMB. The dark ages can be probed effectively using the redshifted \HI 21-cm signal, produced when neutral hydrogen atoms absorb or emit photons at a (local) frequency of 1420\,MHz. The 21-cm signal can be observed over a range of cosmic times, as each frequency corresponds to a different look-back time. This means that the 21-cm signal can be used to map the evolution of the Universe over time. This signal is also naturally three-dimensional (3D), meaning that it can in principle be used to measure the full spatial distribution of neutral hydrogen. This is in contrast to the CMB, which is a 2D signal. The dark ages can be probed by measuring the global (or mean) signal as well as by measuring the 3D power spectrum over the relevant redshift range. Therefore, the 21-cm signal from the dark ages contains in principle more cosmological information than the CMB \citep{Loeb2004, Mondal2023}.

The 21-cm power spectrum during the dark ages is sensitive to the $\Lambda$CDM cosmological parameters, particularly on the scale of the baryon acoustic oscillations (BAOs) \citep{barkana2005, Mondal2023}. In addition to the absence of complicating astrophysics, the fluctuations are still rather linear during the dark ages, so modeling and interpreting them is simpler than for probes in the more recent Universe. The 21-cm fluctuations probe fluctuations of the baryon density, peculiar velocity \citep{bharadwaj04,BLlos}, and baryon temperature \citep{naoz05,barkana2005}. A number of smaller contributions must be included in a precise calculation \citep{Lewis2007, Ali-Ha2014}.

The 21-cm signal from the dark ages is a faint radio signal that can only be detected at very low frequencies, below 45\,MHz. The Earth's ionosphere heavily distorts and eventually blocks radio waves at these frequencies. This means that it is impossible to observe the 21-cm signal from the dark ages using radio telescopes on Earth. To overcome this challenge, scientists are developing lunar and space-based experiments to observe the 21-cm signal from the dark ages. These experiments are being rapidly developed as part of the international race to return to the moon. Some such experiments include: NCLE\footnote{\url{https://doi.org/10.1126/science.aau2004}} (Netherlands-China), DAPPER (USA) \citep{dapper}, FARSIDE (USA) \citep{farside}, DSL\footnote{\url{https://www.astron.nl/dsl2015}} (China), FarView (USA) \citep{dapper}, SEAMS (India) \citep{seams}, PRATUSH\footnote{\url{https://wwws.rri.res.in/DISTORTION/pratush.html}} (India), LuSee Night (USA) \citep{luseenight}, ALO\footnote{\url{https://www.astron.nl/dailyimage/main.php?date=20220131}} (Europe), and ROLSES\footnote{\url{https://www.colorado.edu/ness/ness-projects}} (USA). We note that going to the Moon could present substantial practical advantages beyond just avoiding the Earth’s ionosphere: this would offer a potentially benign environment that is extremely dry and stable and could block out terrestrial radiofrequency interference (on the lunar farside).

Given the great potential for precision cosmology and the rapid observational developments, we recently \citep{Mondal2023} studied the use of the 21-cm signal (both the global signal and power spectrum) during the dark ages to constrain cosmological parameters. We showed that measuring the global 21-cm signal for 1,000 hours would allow us to measure a combination of cosmological parameters to within 10\% accuracy. A longer integration time would improve this accuracy even further, potentially down to 1\% or even better, with a measurement of the cosmic Helium fraction that could exceed CMB measurements by the Planck satellite \citep{Planck:2018}. It would be significantly harder to achieve precision cosmology with 21-cm fluctuations, as it would require a large collecting area of order 10\,km$^2$. This is much larger than the collecting area of any existing radio telescope, but it is possible for future instruments. With 10\,km$^2$, we would be able to achieve a measurement accuracy that is twice as good as the global case (with a 1,000 hour integration for both). Increasing the collecting area of integration time further could eventually beat the Planck accuracy in some cosmological parameter combinations, the Helium fraction, and the total mass of neutrinos. Thus, if we assume standard cosmology, 21-cm observations from the dark ages could potentially lead to major advances in our understanding of the Universe.

Whenever considering a new window on the Universe, it is important to also anticipate the possible discovery space that lies beyond
just standard cosmology. Indeed, there may be exotic (non-standard) models that are allowed by other astrophysical probes and could be detected with 21-cm observations of the dark ages. In general, given the array of observational constraints on cosmic history and on the properties of dark matter, it is not obvious how to construct such models. Fortunately, a number of such models were stimulated by the possible EDGES detection of a strong 21-cm signal during cosmic dawn \citep{Bowman:2018}. While disputed at 95\% significance by the SARAS experiment \citep{SARAS}, with further measurements expected to resolve this tension, the tentative EDGES signal has inspired theories that can be probed over a wide range of possible parameters, independently of whether EDGES turns out to be correct. 

Specifically, the anomalously strong EDGES trough has two main categories of explanations. One category of explanation is the presence of an excess radio background (ERB) at high redshifts, with an intensity significantly higher than the CMB \citep{Bowman:2018, Feng:2018, EwallWice:2018, fialkov19, Mirocha:2019, ewall20}. One possibility for such an ERB is an astrophysically-produced radio background, from sources such as active galactic nuclei \citep[AGN,][]{Biermann:2014, bolgar18, EwallWice:2018, ewall20, Mebane2020},  or star-forming galaxies \citep{condon92, jana2019, Reis2020} at high redshift. However, in order to have an effect on the signal from the dark ages (as opposed to only cosmic dawn), the ERB must have been formed in the early Universe, during or before the dark ages. In \citet{fialkov19} we showed that the depth and steepness of the EDGES signal could be explained by such a homogeneous ERB with a synchrotron spectrum. Exotic processes such as dark matter annihilation or superconducting cosmic strings \citep{Fraser:2018, Pospelov:2018, Brandenberger:2019} could give rise to this kind of homogeneous early ERB. Models of an ERB are also motivated by the observation at low frequencies of an excess radio background over that CMB by ARCADE2 \citep{Fixsen:2011, seiffert11}, confirmed by LWA1 \citep{Dowell:2018} in the frequency range $40-80$ MHz. This observed excess radio may be extragalactic, but it is unclear what fraction of the observed excess originates from Galactic compared to extragalactic sources \citep[e.g.,][]{Subrahmanyan:2013}. In any case, the observed excess serves as an upper limit for an extragalactic ERB. 

The second category of explanations for EDGES is that an additional cooling mechanism cooled the gas faster than just adiabatic cooling due to the cosmic expansion. An additional cooling mechanism has been suggested \citep{Barkana:2018, Berlin, Barkana2018a, munoz18, Liu19} that involves a non-gravitational interaction between the ordinary matter and the dark matter particles (e.g., via Rutherford-like scattering); this drives down the temperature of the gas leading to the strong observed absorption. 

In this paper we consider these two categories, first an homogeneous ERB with a synchrotron spectrum \citep{fialkov19} (where this spectrum is motivated by the possibility that the ERB explains part or all of the observed extragalactic radio background), and then the millicharged dark matter (mDM) model \citep{munoz18} in which a small fraction of the dark matter particles have a tiny electric charge. Throughout the paper, we assume the Planck+BAO best fit values of cosmological parameters \citep[][table 2, last column]{Planck:2018}.

\section{The 21-cm signal}

\begin{figure}
\centering
\includegraphics[width=.48\textwidth, angle=0]{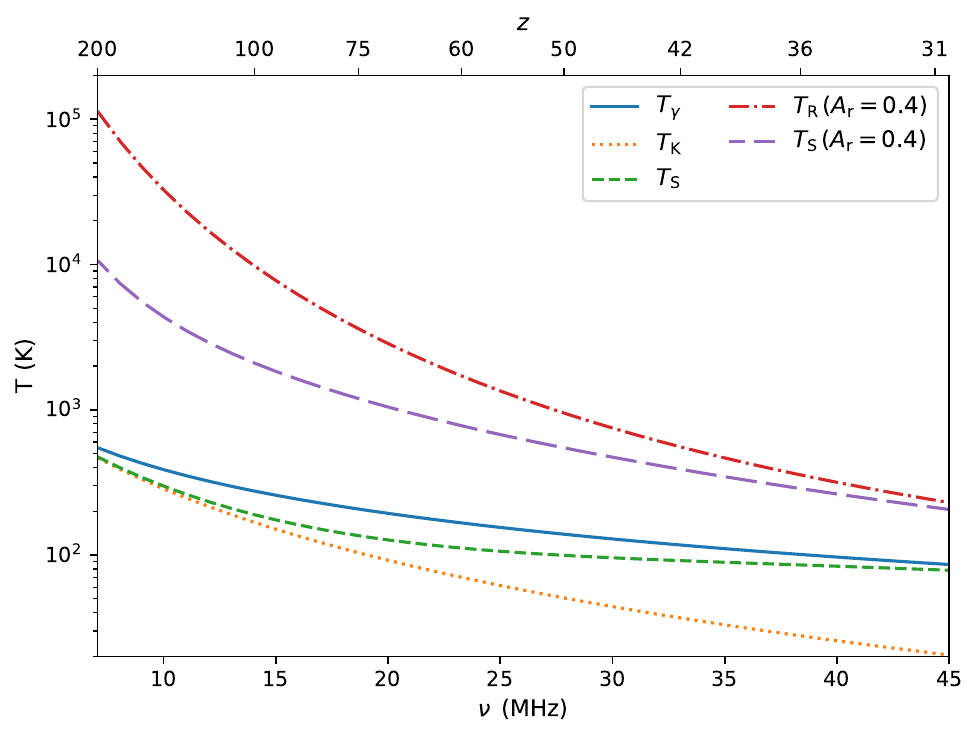}
\caption{The evolution of the CMB temperature $T_{\rm \gamma}$, gas temperature $T_{\rm K}$ and the spin temperature $T_{\rm S}$ (in units of K), as a function of $\nu$ (or $z$ as the top $x$-axis). The total radio background temperature $T_{\rm R}$ for the ERB model with $A_{\rm r} = 0.4$ is also shown, along with the corresponding $T_{\rm S}$.}
\label{fig:Temp_nu}
\end{figure}

The brightness temperature of the 21-cm line, relative to the CMB, is given by the following equation:
\begin{equation}
T_{\rm b} = (T_{\rm S} - T_{\gamma})\frac{1 - e^{-\tau_{21}}}{1 + z} \ ,
\end{equation}
where $T_\gamma = 2.725~{\rm K}~(1+z)$.
Assuming that the optical depth of the 21-cm transition $\tau_{21} \ll 1$, this can be written as:
\begin{multline}
T_{\rm b} = 54.0\,{\rm mK}\, \frac{\rho_{\rm HI}}{\bar{\rho}_{\rm H}} \left(\frac{\Omega_{\rm b}h^2}{0.02242}\right) \left(\frac{\Omega_{\rm m}h^2}{0.1424}\right)^{-\frac{1}{2}} \left(\frac{1 + z}{40}\right)^{\frac{1}{2}} \\
\frac{x_{\rm c}}{1 + x_{\rm c}} \left(1 - \frac{T_{\gamma}}{T_{\rm K}}\right)\ ,
\label{eq:Tb}
\end{multline}
where $\rho_{\rm HI}$ is the neutral hydrogen density, $\bar{\rho}_{\rm H}$ is the cosmic mean density of hydrogen, and $x_{\rm c}$ is the collisional coupling coefficient \citep{Zygelman2005}. In Fig.~\ref{fig:Temp_nu} we show the evolution of $T_{\gamma}$ and $T_{\rm K}$ (cyan-blue and light-orange lines, respectively) as a function of $\nu$ (and $z$). Note that $T_{\gamma}$ falls as $1/\nu$ (or $[1+z]$), while $T_{\rm K}$ falls faster, eventually (at the lower redshifts) as $1/\nu^2$ (or $[1+z]^2$).

\begin{figure}
\centering
\includegraphics[width=.48\textwidth, angle=0]{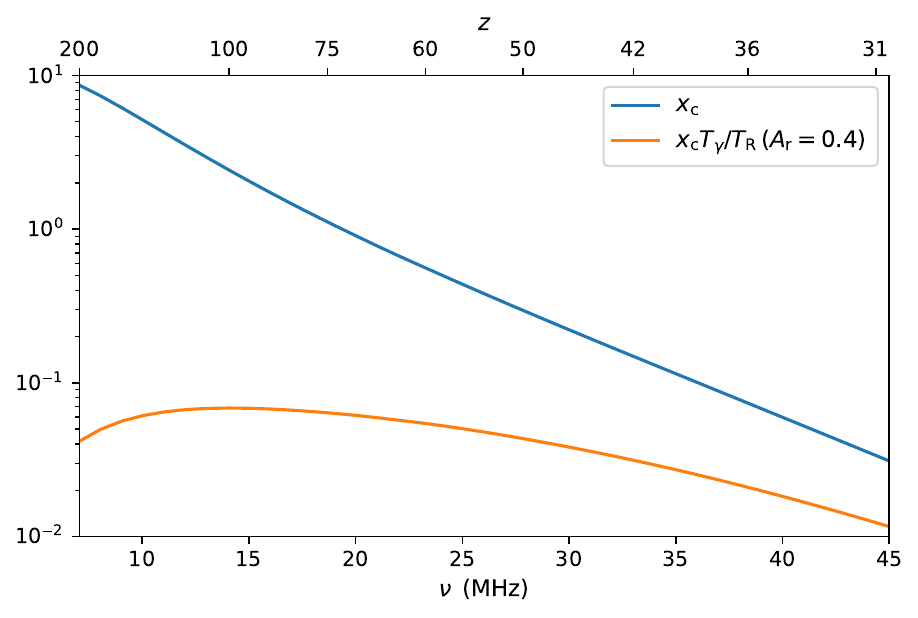}
\caption{The evolution of the collisional coupling coefficient $x_{\rm c}$ as a function of $\nu$ (or $z$ as the top $x$-axis). This also shows $x_{\rm c} T_{\gamma}/T_{\rm R}$, which gives the effective coupling in the case with $T_{\rm R}$ for $A_{\rm r} = 0.4$.}
\label{fig:xc}
\end{figure}

During the dark ages, the spin temperature of the hydrogen atoms is pulled towards the temperature of the gas ($T_{\rm S} \xrightarrow[]{} T_{\rm K}$) by atomic collisions, while it is pulled towards the temperature of the CMB ($T_{\rm S} \xrightarrow[]{} T_{\gamma}$) by CMB scattering. The relative importance of these two effects depends on the density of the gas. Fig.~\ref{fig:xc} shows the evolution of $x_{\rm c}$ as a function of $\nu$ (or $z$). The value of $x_{\rm c}$ is a measure of the efficiency with which collisions between hydrogen atoms can couple the spin states of the \HI atoms into equilibrium with the regular (kinetic) gas temperature. The coupling is strong (and so $T_{\rm S} \approx T_{\rm K}$) roughly until the value $x_{\rm c}=1$ is reached at $z=72.4$, after which $T_{\rm S}$ (which is shown in Fig.~\ref{fig:Temp_nu}) begins to approach the CMB temperature.

The sky-averaged 21-cm brightness temperature, as a function of $\nu$ (or $z$), is referred to the 21-cm global (or mean) signal. Experiments measuring the global signal require a single, well-calibrated antenna. Therefore, they are relatively simple and advantageous to consider as the first step toward detecting the dark ages signal. For the standard set of cosmological parameters, we use the \texttt{CAMB}\footnote{\url{http://camb.info}} \citep{Lewis2007,CAMB} cosmological perturbation code to precisely generate the 21-cm global signal\footnote{To extract the 21-cm global signal from \texttt{CAMB}, we run \texttt{CAMB} twice, once with temperature units on and once with temperature units off, and take the ratio of the transfer functions in the two cases.}. The 21-cm global signal during the dark ages is always negative (corresponding to absorption relative to the CMB). The 21-cm global signal from the dark ages in the standard cosmological model is shown in comparison with other cases later in this paper (e.g., the black dashed line in Fig.~\ref{fig:global_radio}). The peak of the signal is 40.2\,mK at $\nu = 16.3$\,MHz ($z = 86$).

In addition, the dark ages can be probed by measuring the fluctuations in the 21-cm signal at various length scales, i.e., the power spectrum. These fluctuations are mainly due to the fluctuations in the gas density, temperature, and $x_{\rm c}$. To accurately predict the 21-cm power spectrum, we use \texttt{CAMB} [which includes small additional effects \citep{Lewis2007, Ali-Ha2014} not included in eq.~(\ref{eq:Tb})] and add to it redshift space distortions caused by the line-of-sight component of the peculiar velocity of the gas \citep{kaiser87, bharadwaj04, BLlos} and the light-cone effect \citep{barkana06,mondal18}, as detailed in our previous paper \citep{Mondal2023} [Note that the Alcock-Paczy\'{n}ski effect \citep{AP1979,AliAP,Nusser,APeffect} is not relevant since we do not vary the cosmological parameters in this paper]. The 21-cm power spectrum from the dark ages for the standard cosmological model is shown in comparison with other cases later in this paper (e.g., the dashed lines in Figs.~\ref{fig:power} and \ref{fig:power_nu}). The power increases initially as the adiabatic expansion cools the gas faster than the CMB, and density fluctuations grow due to gravity. However, eventually the power decreases as the declining density reduces $x_{\rm c}$. For example, the maximum squared fluctuation $\Delta^2$ at $k = 0.1$\,Mpc$^{-1}$ is 0.44\,mK$^2$ at $z = 51$. Measuring the dark ages power spectrum is substantially more difficult than measuring the global signal, but it contains potentially much more information \citep{Loeb2004}. As we have recently shown \citep{Mondal2023}, for standard cosmology the global signal offers a relatively accessible first step to observing the dark ages, with the power spectrum requiring a much greater investment to get started, but offering far greater potential returns. More specifically, a single lunar global antenna can make a novel test of the standard cosmological model, showing whether it can describe the dark ages or if instead there is some surprise in cosmic history. An array of antennas (either global antennas for increased integration time, or an interferometric array) can yield some cosmological parameters (the overall baryon density and the Helium fraction) at an accuracy competitive with Planck, and a very large interferometer can outperform Planck on these parameters as well as the total mass of neutrinos.

\section{The excess radio background model}

\subsection{Global signal}

To calculate the global signal for the excess radio background (ERB) model, we use eq.~(\ref{eq:Tb}) together with the value of $T_{\rm b}$ from \texttt{CAMB}, in order to extract $x_c$ (which we show in Fig.~\ref{fig:xc}; note that this calculation neglects the residual ionized fraction and other tiny effects). Now, in the presence of a radio background, we change the final factors in eq.~(\ref{eq:Tb}) from
$$ 
\frac{x_{\rm c}}{1 + x_{\rm c}} \left(1 - \frac{T_{\gamma}}{T_{\rm K}}\right) 
$$ 
to
$$
\frac{x_{\rm c} T_{\gamma}/T_{\rm R}}{ 1 + x_{\rm c} T_{\gamma}/T_{\rm R}} \left(1 - \frac{T_{\rm R}}{T_{\rm K}}\right)\ .
$$
Here, $x_{\rm c}$ is the same value as before, i.e., we use it to denote the value in the absence of the ERB; note that this differs from the notation used in some previous papers. The effective coupling constant in the ERB case is $x_{\rm c} T_{\gamma}/T_{\rm R}$ (which has often been denoted just $x_{\rm c}$ in the ERB case). We use this notation in order to show simply and clearly how the ERB changes the 21-cm signal (as opposed to the previous notation which hides part of the ERB effect within the change in $x_{\rm c}$). We note that an additional CMB heating mechanism suggested by \citet{venumadhav18} would be even more important in the presence of an ERB; however, \citet{Meiksin21} showed that this mechanism corresponds to a balanced internal energy exchange that does not significantly heat the gas and thus should not be included. 

The total radio background at 21 cm at redshift $z$ (including the CMB plus the ERB) is assumed to be (as in \citealt{fialkov19}):
\begin{equation}
T_{\rm R} = T_{\gamma} \left[ 1 + A_{\rm r} \left( \frac{\nu_{\rm obs}} {78\ {\rm MHz}} \right)^{\alpha}\right]\ ,
\end{equation}
where $\nu_{\rm obs}  = 1420\ {\rm MHz}/(1+z)$. Here the amplitude $A_{\rm r}$ of the ERB is measured relative to the CMB at an observed frequency of 78~MHz, approximately the center of the tentative EDGES absorption feature. We assume $\alpha = -2.6$ to match the spectrum of the extragalactic radio background. \citet{fialkov19} showed that a minimum value of $A_{\rm r}=1.9$ is required in order to match the EDGES feature, when combined with models covering a wide range of possible astrophysical parameters. The level of the extragalactic radio background (its 2$\sigma$ upper limit) gives an upper limit of 375 on the possible value of $A_{\rm r}$. 

\begin{figure}
\centering
\includegraphics[width=.48\textwidth, angle=0]{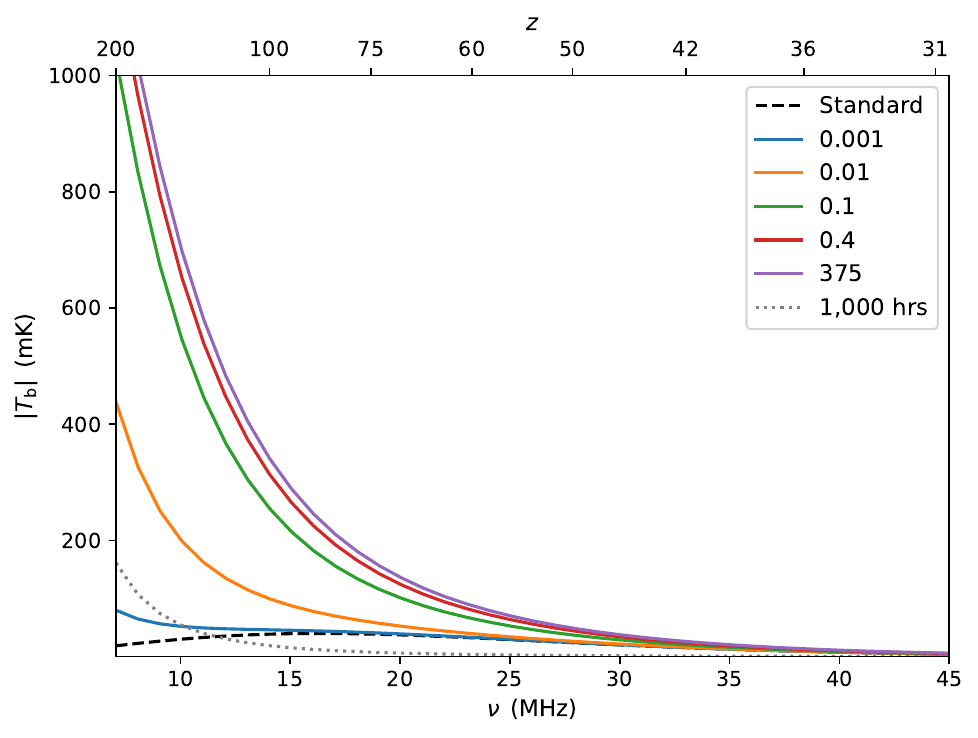}
\includegraphics[width=.48\textwidth, angle=0]{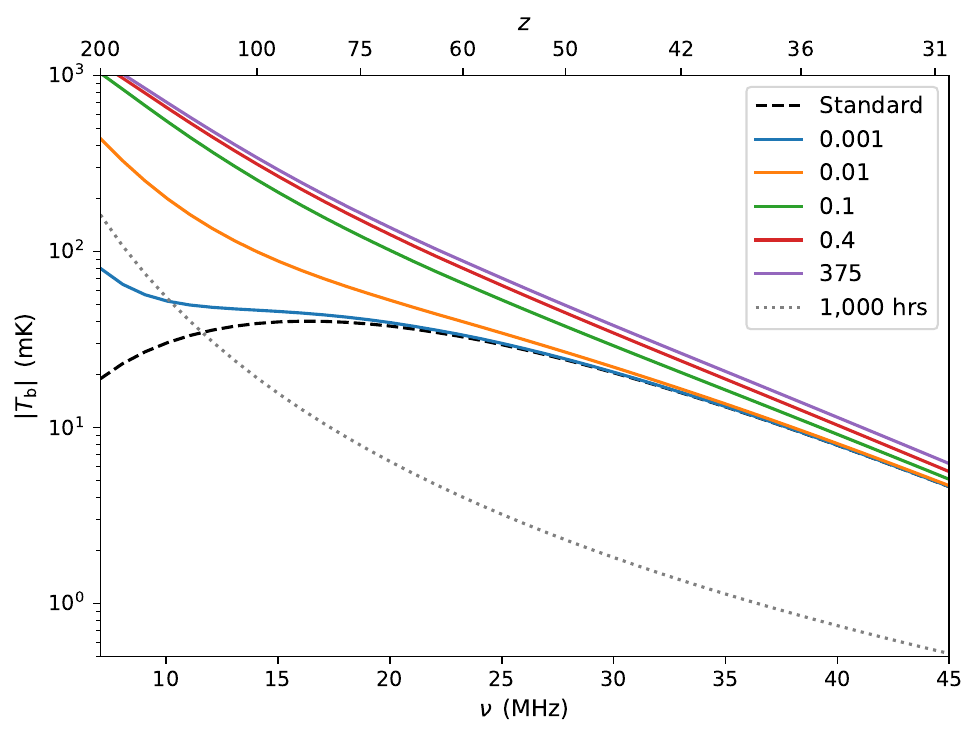}
\caption{The size of the global 21-cm signal as a function of $\nu$ for the standard $\Lambda$CDM model (black dashed line) and the excess radio model (solid lines) with $A_{\rm r} = 0.001$, 0.01, 0.1, 0.4 and 375. We also show the expected thermal noise for a global signal experiment observing for integration time 1,000 hrs (grey dotted line) for a bin around each $\nu$ of width $\Delta(\ln \nu) = 1$. The same results are shown twice, with either a standard linear $y$-axis (top panel) or a logarithmic $y$-axis (bottom panel) for an easier comparison among the models.}
\label{fig:global_radio}
\end{figure}

Fig.~\ref{fig:global_radio} shows the size of the global 21-cm signal from the dark ages as a function of $\nu$ (and $z$), for the excess radio background model with various values of $A_{\rm r} = [0.001$, 0.01, 0.1, 0.4, 375]. Also shown is the standard case which corresponds to $A_{\rm r} = 0$, i.e., CMB-only and no ERB. The absorption signal increases sharply with $z$ for the ERB models (hence the $y$-axis of the bottom panel is logarithmic, which is unusual for plots of the global signal). The signal also increases with $A_{\rm r}$, but even $A_{\rm r} = 0.1$ nearly saturates the dark ages signal, i.e., the signal becomes independent of $A_{\rm r}$ only slightly beyond that value. In Fig.~\ref{fig:global_radio}, we also show the instrumental noise for integration time $t_{\rm int} = 1$,000\,hrs for a bin around each $\nu$ of width $\Delta(\ln \nu) = 1$. 
This lets us illustrate the overall signal-to-noise ratio (S/N) in the figure, using a bin size of order the central value. For the Fisher matrix predictions, though, we used 40 frequency (or redshift) bins in the range $6.56 \le \nu \le 46.56$, with a bin width of $\Delta \nu = 1\,{\rm MHz}$. We chose the upper end of the frequency range to be $z \approx 30$, which is the typical redshift where galaxies at cosmic dawn first form in sufficient numbers to significantly affect the 21-cm signal \citep{subtle}.
As expected, the noise increases sharply with redshift. Indeed, the redshift dependence of the thermal noise and of the saturated radio signal are somewhat similar, by coincidence (see below).

\begin{figure}
\centering
\includegraphics[width=.48\textwidth, angle=0]{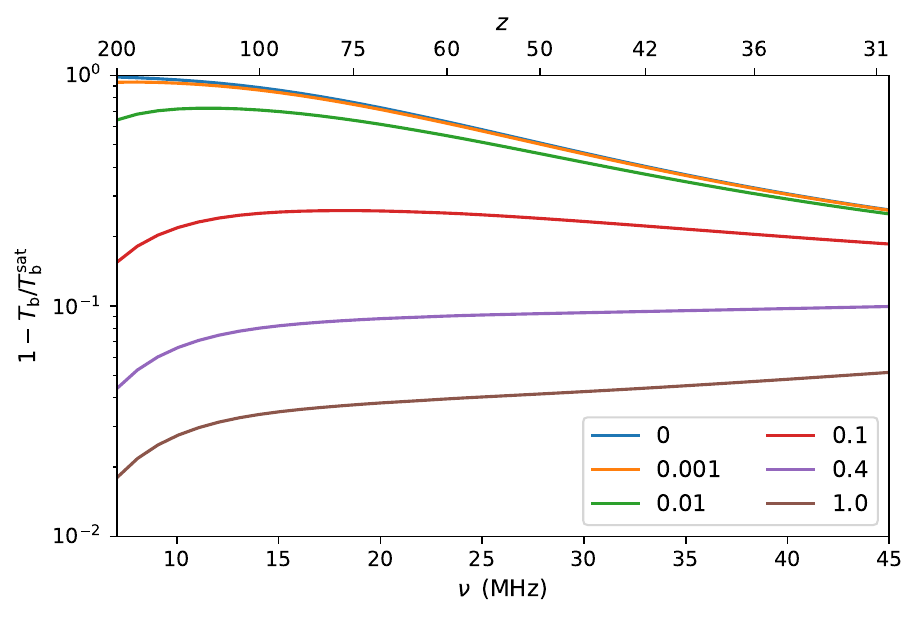}
\caption{The fractional difference $[1-T_{\rm b}(A_{\rm r})/T_{\rm b}^{\rm sat}]$ as a function of $\nu$ for $A_{\rm r} = 0$,\,0.001, 0.01, 0.1, 0.4 and 1. Here $T_{\rm b}^{\rm sat}$ is the global signal for the case $A_{\rm r} = 375$, and the case $A_{\rm r} = 0.4$ corresponds to $A_{\rm r90}$ (at least 90\% saturation throughout the dark ages).}
\label{fig:Tb_frac}
\end{figure}

We now label as the saturated ERB signal the case with the maximum $A_{\rm r} = 375$, and refer to the 21-cm brightness temperature in this case $T_{\rm b}^{\rm sat}$. We then examine the approach to saturation by showing the fractional difference $[1-T_{\rm b}(A_{\rm r})/T_{\rm b}^{\rm sat}]$, which is always positive, as a function of $\nu$. Fig.~\ref{fig:Tb_frac} shows this for $A_{\rm r}=0$, 0.001, 0.01, 0.1, 0.4, and 1. We find that the maximum value of the fractional difference is 0.1 for $A_{\rm r}=0.4$, so that this is the value that gives at least 90\% saturation throughout the dark ages. Thus, it is a minimum $A_{\rm r}$ value for being near saturation, which we label $A_{\rm r90}$.

Before continuing, we use the $A_{\rm r}=0.4$ case to illustrate how the ERB affects the 21-cm signal. Fig.~\ref{fig:Temp_nu} shows $T_{\rm S}$ for this ERB case, and Fig.~\ref{fig:xc} shows the effective collisional coupling coefficient in the same case. On the one hand, $T_{\rm R}$ in this ERB case is higher than $T_{\rm \gamma}$, with a relative factor that rises rapidly towards high redshift. On the other hand, the effective coupling is suppressed by the high radio background, which makes it actually decrease with redshift at the high end. As a result, the effective coupling coefficient never even reaches as high as 0.1. The balance between the high radio background and the low effective coupling keeps $T_{\rm S}$ from coming too close to $T_{\rm R}$ at high redshifts, and leads to a saturated signal (in the limit of $T_{\rm R} \rightarrow \infty$, or more specifically $T_{\rm R} \gg x_{\rm c} T_{\gamma}$ and also $T_{\rm R} \gg T_{\rm K}$) with a value of 
\begin{equation}
T_{\rm b}^{\rm sat} = - 54.0\,{\rm mK}\, \left(\frac{\Omega_{\rm b}h^2}{0.02242}\right) \left(\frac{\Omega_{\rm m}h^2}{0.1424}\right)^{-\frac{1}{2}} \left(\frac{1 + z}{40}\right)^{\frac{1}{2}} 
x_{\rm c} \frac{T_{\gamma}}{T_{\rm K}}\ ,
\end{equation}
consistent with the dark ages section in \citet{fialkov19}. This global signal turns out to have a spectral shape that is somewhat similar to the foreground, which induces a partial degeneracy between them (see below). We emphasize that this is a coincidence. Indeed, the saturated signal spectrum is unrelated to the ERB spectrum, and is driven mainly by the collisional coupling coefficient $x_{\rm c}$, which rises roughly as $(1+z)^3$ (driven largely by the increasing cosmic density).

When we consider the ability of lunar or space-based global experiments to measure the dark ages signal, we account for foreground removal (in an optimistic scenario) by adding a term in the shape of the synchrotron foreground (i.e., $A \nu^{-2.6}$ with the amplitude a free parameter). Here, the sky brightness temperature $T_{\rm sky} = 180\times (\nu/180\,{\rm MHz})^{-2.6}$\,K \citep{Furlanetto2006}. A contributing component of this shape in the model cannot be distinguished from the foreground. Then, for any signal, we can determine the statistical significance of its detection (i.e., the detection of the difference between the expected signal and a zero signal) as follows. We define the signal as a parameter $\beta$ times the expected signal (i.e., the expected signal corresponds to $\beta$=1, and the absence of the signal to $\beta=0$. We then fit to the data, using a Fisher analysis to extract the error $\delta \beta$ in the measurement of $\beta$ (assuming all cosmological parameters are fixed at their fiducial values as determined by Planck). This tells us the significance of the detection of the signal relative to zero, i.e., assuming Gaussian thermal noise, the detection significance is a number of $\sigma$ equal to $1/\delta \beta$. For the estimation of noise in the global signal measurement, we assume a redshift range of 30-200 with a bandwidth of $\Delta \nu=1$\,MHz and explore three different integration times of $t_{\rm int} = 1$,000\,hrs, 10,000\,hrs and 100,000\,hrs. We note that in practice, an array of global antennas can be used to increase the total effective integration time. 

We first consider the significance of the detection of the standard global signal, without an ERB. We find that is would be distinguishable from zero at $4.12 \sigma$ for $t_{\rm int}$ of 1,000\,hrs, and $41.2 \sigma$ for 100,000\,hrs (see the upper panel of Table~\ref{tab:standard}).

\begin{table}
\begin{center}
\caption{The significance (\# of $\sigma$) of the detection of the standard signal.}
\label{tab:standard}
\begin{minipage}{0.48\textwidth}

\begin{tabular*}
{\textwidth}{@{\extracolsep{\fill}}lccc@{\extracolsep{\fill}}}
\hline

 & \multicolumn{3}{@{}c@{}}{Integration time} \\
\hline
 & 1,000\,hrs & 10,000\,hrs & 100,000\,hrs \\

\hline
Global signal & 4.12 & 13.0 & 41.2
\end{tabular*}
\end{minipage}
\begin{minipage}{0.48\textwidth}

\begin{tabular*}
{\textwidth}{@{\extracolsep{\fill}}lccccc@{\extracolsep{\fill}}}
\hline

 & \multicolumn{5}{@{}c@{}}{Configuration} \\
\hline
 & G & A & B & C & D \\

\hline
Power spectrum & 3.01 & 6.71 & 66.6 & 81.6 & 690 \\
\hline
\end{tabular*}
\end{minipage}
\end{center}
\end{table}

\begin{figure}
\centering
\includegraphics[width=.48\textwidth, angle=0]{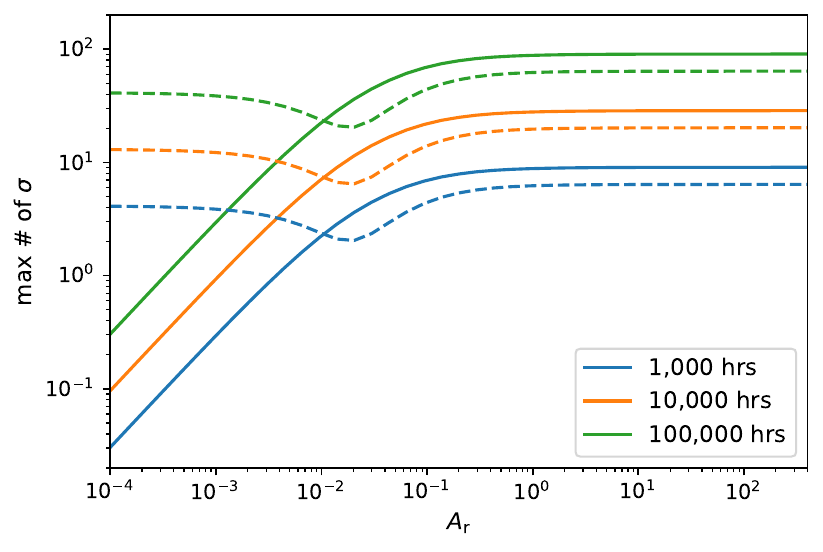}
\caption{The significance (\# of $\sigma$) of the detection (i.e., $1/\delta \beta$) as a function of $A_{\rm r}$ (minimized over $z_{\rm r}$; see the text). We show two detection scenarios: distinguishing the ERB global signal from the standard case (solid lines), and detecting it relative to zero signal (dashed lines).}
\label{fig:dbeta_Ar}
\end{figure}

For the ERB models, there are two interesting cases to ask: how well they can be detected (i.e., distinguished from a zero signal), and how well they can be distinguished from the standard signal (i.e., showing that the signal is anomalous and must correspond to exotic physics); in the latter case, the signal model corresponds to the standard signal plus $\beta$ times the difference between the ERB model and the standard model. Fig.~\ref{fig:dbeta_Ar} shows the significance of the detection (i.e., $1/\delta \beta$) of the signal relative to the standard signal (the solid curves), and the significance of a detection in general (i.e., relative to zero; dashed curves). Depending on the value of $A_{\rm r}$, either significance can be higher. The significance of the detection relative to the standard signal increases with increasing $A_{\rm r}$ (as the ERB signal differs more and more from the standard case), initially going as $A_{\rm r}$ until it saturates roughly beyond $A_{\rm r90}$. The significance of the detection relative to zero signal goes between that for the standard signal (at small $A_{\rm r}$) to that for the saturated ERB signal (at large $A_{\rm r}$), with a small trough at $A_{\rm r} = 0.0204$; this is the value of $A_{\rm r}$ with the smallest significance of detecting the ERB signal (relative to zero signal), the significance being 2.04$\sigma$, 6.46$\sigma$, and 20.4$\sigma$, for $t_{\rm int}= 1$,000\,hrs, 10,000\,hrs, and 100,000\,hrs, respectively.

The significance of detecting the ERB global signal in these two ways is also listed in Table~\ref{tab:ERB}, for various values of $A_{\rm r}$ and $t_{\rm int}$. A 1,000\,hr global experiment can detect the saturated ERB signal at 6.38$\sigma$ significance, and distinguish it from the standard signal at 9.04$\sigma$. We note that both of these are substantially stronger statistical results than the detection of the standard signal itself ($4.12 \sigma$ in this case), due to the greater amplitude of the signal in the ERB case. With $t_{\rm int}= 100$,000\,hrs, the significance levels would increase $\times$10. 

\begin{table}
\begin{center}
\caption{The significance (\# of $\sigma$) of detecting the ERB global signal relative to zero signal or the standard signal.}
\label{tab:ERB}
\begin{minipage}{0.48\textwidth}

\begin{tabular*}
{\textwidth}{@{\extracolsep{\fill}}lcccc@{\extracolsep{\fill}}}
\hline

 & & \multicolumn{3}{@{}c@{}}{Integration time} \\
\hline
 & $A_{\rm r}$ & 1,000\,hrs & 10,000\,hrs & 100,000\,hrs \\

\hline
 & 0.001 & 3.86 & 12.2 & 38.6 \\
 & 0.01 & 2.38 & 7.50 & 23.7 \\
Relative to zero & 0.1 & 4.40 & 13.9 & 44.0 \\
 & 0.4 & 5.89 & 18.6 & 58.9 \\
 & 375 & 6.38 & 20.2 & 63.8 \\
\hline
 & 0.001 & 0.292 & 0.922 & 2.92 \\
 & 0.01 & 2.22 & 7.02 & 22.2 \\
Relative to standard & 0.1 & 6.91 & 21.8 & 69.1 \\
 & 0.4 & 8.45 & 26.7 & 84.5 \\
 & 375 & 9.04 & 28.6 & 90.4 \\
\hline
\end{tabular*}
\end{minipage}
\end{center}
\end{table}

\begin{figure}
\centering
\includegraphics[width=.48\textwidth, angle=0]{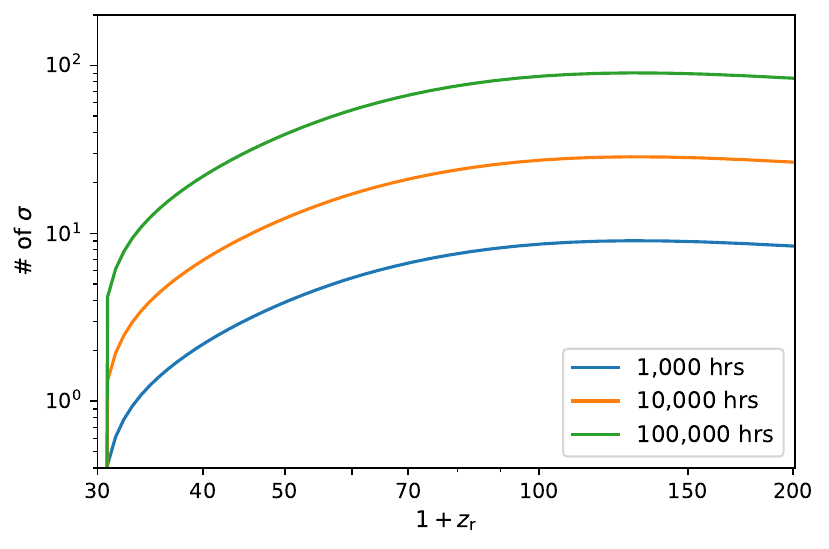}
\caption{The significance (\# of $\sigma$) of the detection (i.e., $1/\delta \beta$) for distinguishing the ERB global signal from the standard case, shown as a function of the maximum redshift $z_{\rm r}$, for the saturated $A_{\rm r} = 375$ case.}
\label{fig:dbeta_zr}
\end{figure}

There are some subtleties in these results. First, there is the issue of how it is possible that distinguishing the ERB signal from the standard signal is easier than from zero, in some cases. The answer is the degeneracy with the foreground term; since the ERB signal (most clearly in the saturated case) has a shape versus frequency that is similar to the foreground, it is more difficult to detect it than would be expected just based on its amplitude, while the standard global signal has a shape that differs more clearly from the shape of the foreground. A related subtlety has to do with the redshift range of the fitting. When measuring the ERB signal relative to the standard signal, including the highest redshifts adds more information, but this goes into determining more accurately the foreground term (i.e., the signal component that is degenerate with the foreground) rather than the ERB signal itself; in other words, $\delta \beta$ is higher due to the stronger degeneracy with the foreground term. Fig.~\ref{fig:dbeta_zr} shows the significance of the detection (for ERB detection relative to the standard model) versus the maximum redshift $z_{\rm r}$, for the saturated case ($A_{\rm r} = 375$). The maximum for all the curves occurs at $z_{\rm r} = 126$. For example, for $t_{\rm int} = 1$,000\,hrs, the maximum significance is $9.0\,\sigma$ while the value for $z_{\rm r} = 200$ is $8.4\,\sigma$. Thus, actually Fig.~\ref{fig:dbeta_Ar} shows the significance not for $z_{\rm r}=200$ but rather for the value of $z_{\rm r}$ that gives the maximum significance in each case. We note that Fig.~\ref{fig:dbeta_zr} also doubles as showing the case when the excess background was produced only at redshift $z_{\rm r}$ (and not before), and we fit to observations only up to that $z_{\rm r}$. For example, a measurement of the dark ages global 21-cm signal between redshifts 30 and 56 to the precision of thermal noise from a 1,000\,hour integration would be able to distinguish the saturated signal from the standard signal at $5\sigma$. 

Looking again at Fig.~\ref{fig:dbeta_Ar}, we note that the significance for detecting the signal (relative to zero) varies over a limited range. However, the significance for distinguishing the ERB model from the standard cosmological model varies over a wide range, which leads to the question of how small a value of $A_{\rm r}$ can be distinguished. Table~\ref{tab:sig_para} shows the $A_{\rm r}$ values for which the ERB model can be distinguished from the standard model at various confidence levels. For example, a measurement of the global 21-cm signal to the precision of thermal noise from a 1,000\,hour integration would be able to detect a minimum value of $A_{\rm r} = 0.0389$ at $5\sigma$. This is lower than the minimum value required to explain EDGES by a factor of 49; it also corresponds to only $0.0104\%$ of the value that would explain the entire extragalactic radio background (what we took as the saturated ERB case). A 100,000\,hour integration would detect an $A_{\rm r}$ as low as 0.00175 at $5\sigma$.

\begin{table}
\begin{center}
\caption{The minimum value of $A_{\rm r}$ (in units of $10^{-2}$) that allows the ERB global signal to be distinguished from the standard ($A_{\rm r}=0$) case, at various levels of statistical significance.}
\label{tab:sig_para}

\begin{minipage}{0.48\textwidth}

\begin{tabular*}
{\textwidth}{@{\extracolsep{\fill}}lccccc@{\extracolsep{\fill}}} 
\hline

& & \multicolumn{4}{@{}c@{}}{Detection limit} \\
\hline
{} & Integration time & 1$\sigma$ & 2$\sigma$ & 3$\sigma$ & 5$\sigma$ \\

\hline
 & 1,000 hrs & 0.375 & 0.866 & 1.53 & 3.89 \\
$A_{\rm r}\,[10^{-2}]$ & 10,000 hrs & 0.108 & 0.225 & 0.353 & 0.640 \\
 & 100,000 hrs & 0.0333 & 0.0673 & 0.102 & 0.175 \\

\hline
\end{tabular*}
\end{minipage}
\end{center}
\end{table}

\subsection{Power spectrum}
\label{sec:power}
As discussed earlier, measuring the power spectrum is a much more challenging task than measuring the global signal. However, while the global 21-cm signal from the dark ages is a powerful cosmological probe, it is limited in what it can tell us. The power spectrum, on the other hand, is a much richer dataset that has the potential to reveal a wealth of information about the early Universe. To calculate the power spectrum signal for the ERB model, we note that if the brightness temperature is written as a product of various factors, then taking the logarithm and then the derivative shows that, at linear order, the fractional (relative) perturbation in the temperature is the sum of the fractional perturbation in each of the factors. In particular, using eq.~(\ref{eq:Tb}), we get (again neglecting fluctuations in the residual ionized fraction and other tiny effects within CAMB):
\begin{equation}
\delta_{T{\rm b}} = \delta + \frac{1}{1+x_{\rm c}} \delta_{xc}  + \frac{T_{\gamma}}{T_{\rm K} - T_{\gamma}}\delta_{T{\rm K}} \ ,
\label{eq:dtb}
\end{equation}
where each $\delta$ denotes a dimensionless (fractional) perturbations, in the baryon density ($\delta$), the collisional coupling coefficient ($\delta_{xc}$), and the gas temperature ($\delta_{T{\rm K}}$). At each redshift we extract $\delta_{xc}$ from this expression using \texttt{CAMB} to get all the other quantities (including $\delta_{T{\rm b}}$).

Now, in the presence of a radio background, eq.~(\ref{eq:dtb}) becomes instead:
\begin{equation}
\delta_{T{\rm b}} = \delta + \frac{T_{\rm R}}{T_{\rm R}+x_{\rm c} T_{\gamma}} \delta_{xc}  + \frac{T_{\rm R}}{T_{\rm K} - T_{\rm R}}\delta_{T{\rm K}} \ .
\end{equation}
From this we get the monopole 21-cm perturbation in the ERB model. As mentioned in the introduction for the standard case, we then add the effect of line-of-sight velocity gradients, the light-cone effect, and also the effect of the angular resolution of radio interferometers, as in our previous work \citep{Mondal2023}. Henceforth we consider only the total, spherically-averaged power spectrum of 21-cm brightness fluctuations.

\begin{figure}
\centering
\includegraphics[width=.48\textwidth, angle=0]{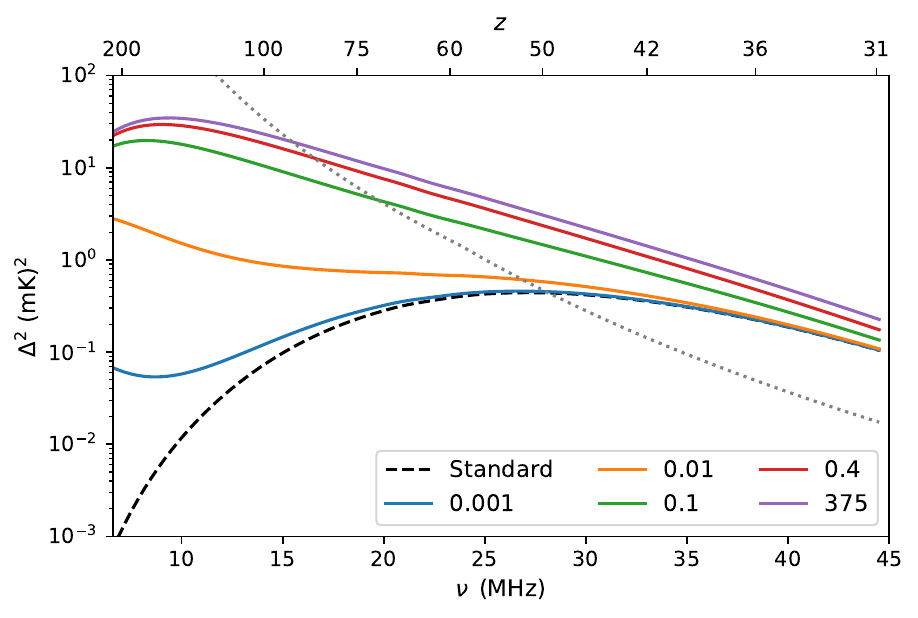}
\caption{The 21-cm power spectrum at $k = 0.1$\,Mpc$^{-1}$ as a function of $\nu$ (or $z$ as the top $x$-axis) for the standard $\Lambda$CDM model (black dashed line) and the excess radio model (solid lines) at various $A_{\rm r}$. We also show the 1$\sigma$ noise (thermal plus cosmic variance) for our G configuration (grey dotted line).}
\label{fig:power_sat}
\end{figure}

Fig.~\ref{fig:power_sat} shows the 21-cm power spectrum as a function of $\nu$ (or $z$), for the standard $\Lambda$CDM model and for the ERB model at various values of $A_{\rm r}$, at $k = 0.1$\,Mpc$^{-1}$. The power spectrum rises with $A_{\rm r}$ until it saturates for $A_{\rm r} \gtrsim A_{\rm r90}$. This is similar to the global case (compare Fig.~\ref{fig:global_radio}), except that the rise is tempered at the highest redshifts (compared to the increasing steepness for the global signal), since the relative fluctuations (in density as well as the other quantities) were smaller at early times. To illustrate plausible measurements of the 21-cm power spectrum from the dark ages, we follow \citet{Mondal2023} and assume five observational configurations, which are listed in Table~\ref{tab:conf}. Here G is meant to roughly correspond to the statistical power of the global case with for $t_{\rm int} = 1$,000\,hrs (in terms of the measurement accuracy of a combination of cosmological parameters; \citealt{Mondal2023}), A doubles the collecting area, B and C increase $\times$10 one of the array parameters from A, and D increases both parameters. In Fig.~\ref{fig:power_sat}, we also show the 1$\sigma$ noise (thermal plus cosmic variance, as in \citet{Mondal2023}) for our minimal G configuration (for bins with $\Delta(\ln \nu) = 1$ and $\Delta(\ln k)=1$); for the power spectrum, the noise increases with redshift significantly faster than the signal, even in the ERB cases.
For the Fisher matrix predictions using the power spectrum, we used 8 frequency (or redshift) bins in the range $5.81 \le \nu \le 45.81$ with a bin width of $\Delta \nu = 5\,{\rm MHz}$ and 11 logarithmic $k$ bins covering the range $0.00779 \le k < 1.91\,{\rm Mpc}^{-1}$ with bin width $\Delta (\ln k) = 0.5$.

\begin{table}
\begin{center}
\caption{The 21-cm power spectrum observational configurations in terms of the collecting area $A_{\rm coll}$ and integration time $t_{\rm int}$.}
\label{tab:conf}
\begin{minipage}{8.5cm}
\begin{tabular*}
{\textwidth}{@{\extracolsep{\fill}}lccccc@{\extracolsep{\fill}}}
\hline

& \multicolumn{5}{@{}c@{}}{Configuration}\\
\hline
{} & D & C & B & A & G\\

\hline

$A_{\rm coll}$ [km$^2$]& 100 & 100 & 10 & 10 & 5 \\
$t_{\rm int}$ [hrs]& 10,000 & 1,000 & 10,000 & 1,000 & 1,000 \\

\hline
\end{tabular*}
\end{minipage}
\end{center}
\end{table}

\begin{figure}
\centering
\includegraphics[width=.48\textwidth, angle=0]{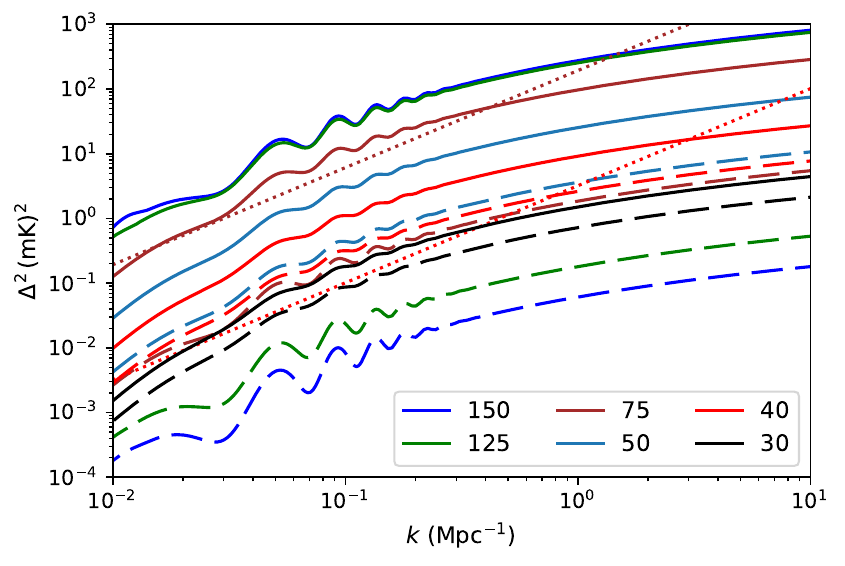}
\caption{The 21-cm power spectrum as a function of wavenumber $k$ during the dark ages, for the standard $\Lambda$CDM model (long dashed lines) and the ERB model (solid lines) with $A_{\rm r} = 375$, at redshifts $z = [150$, 125, 75, 50, 40, 30]. We also show the 1$\sigma$ noise (thermal plus cosmic variance) for our G configuration (dotted lines), at $z = 75$ and 40 (for bins with $\Delta(\ln \nu) = 1$ and $\Delta(\ln k)=1$).}
\label{fig:power}
\end{figure}

Fig.~\ref{fig:power} also shows the 21-cm power spectrum, now as a function of wavenumber $k$ at various redshifts during the dark ages, for the standard $\Lambda$CDM model and for the saturated ERB model with $A_{\rm r} = 375$. The shapes of the power spectra are almost the same for the standard and ERB models (roughly following the shape of the density power spectrum), but the amplitude behaves quite differently. We also show in Fig.~\ref{fig:power} the 1$\sigma$ noise (thermal plus cosmic variance) for the G configuration, at $z = 75$ and 40. As the noise increases rapidly with redshift, and the maximum signal-to-noise ratio (S/N) occurs at the minimum redshift we consider, i.e., $z = 30$.

\begin{figure}
\centering
\includegraphics[width=.48\textwidth, angle=0]{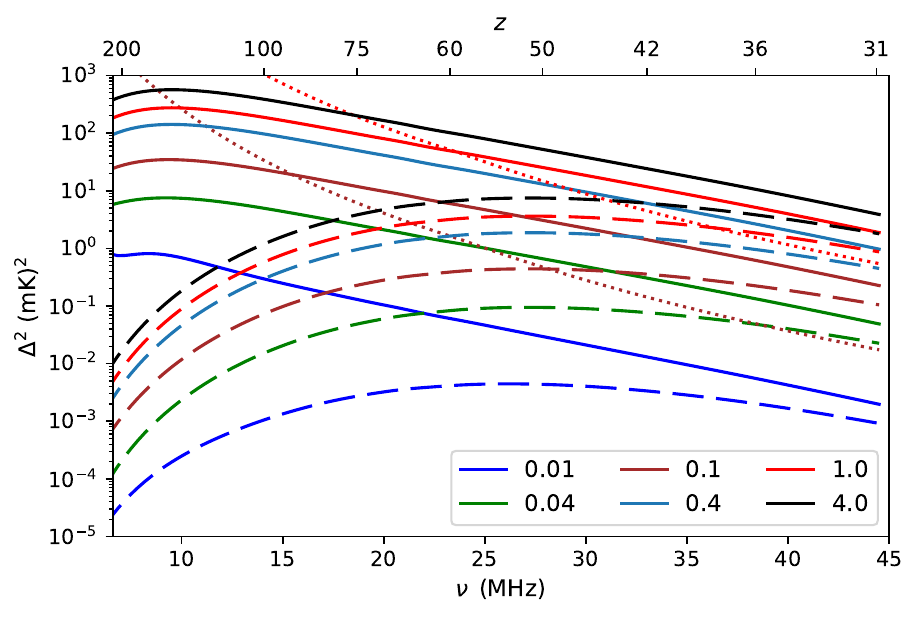}
\caption{The 21-cm power spectrum as a function of $\nu$ (or $z$ as the top $x$-axis) for the standard $\Lambda$CDM model (long dashed lines) and the excess radio model (solid lines) with $A_{\rm r} = 375$, at wavenumber values $k=[0.01$, 0.04, 0.1, 0.4, 1.0, $4.0]\,{\rm Mpc}^{-1}$. We also show the 1$\sigma$ noise (thermal plus cosmic variance) for our G configuration (dotted lines), at $k = 0.1$\,Mpc$^{-1}$ and 1\,Mpc$^{-1}$.}
\label{fig:power_nu}
\end{figure}

Fig.~\ref{fig:power_nu} again shows the 21-cm power spectrum but now in the other cut in terms of the two variables, i.e., as a function of $\nu$ (or $z$) at various wavenumbers, for the standard $\Lambda$CDM model and for the saturated ERB model. Here it is easier to see the difference between the two models. For the standard case, as discussed above, the power spectrum increases initially with time as the amplitude of the signal increases due to the gas cooling faster than the CMB, and as fluctuations increase due to gravity. However, the power spectrum then decreases as the declining density reduces $x_{\rm c}$. In contrast, for the ERB model, the strong radio background at high redshifts results in the power spectrum decreasing monotonically with time over most of the redshift range. As expected, the power spectrum also increases as we go from large scales to small scales. We also show the 1$\sigma$ noise (thermal plus cosmic variance) for the G configuration, at $k = 0.1$\,Mpc$^{-1}$ and 1\,Mpc$^{-1}$. 

Before we consider the power spectrum measurements for the ERB case, as in the global case we first consider the detection significance of the standard power spectrum signal (relative to zero signal). The standard power spectra would be distinguishable from zero at $3.01\sigma$ for the G configuration, going up to $690\sigma$ for configuration D (see the lower panel of Table~\ref{tab:standard}). In terms of the detection significance, each configuration is comparable to a certain integration time for a global experiment: G [534\,hrs], A [2,650\,hrs], B [261,000\,hrs], C [392,000\,hrs], and D [28.0\,million\,hrs]. This demonstrates our conclusion from \citet{Mondal2023} that it is more difficult to start with the 21-cm power spectrum (as even the G configuration requires quite a large collecting area), but eventually interferometers can gather far more cosmological information than is plausible for global experiments. 

\begin{figure}
\centering
\includegraphics[width=.48\textwidth, angle=0]{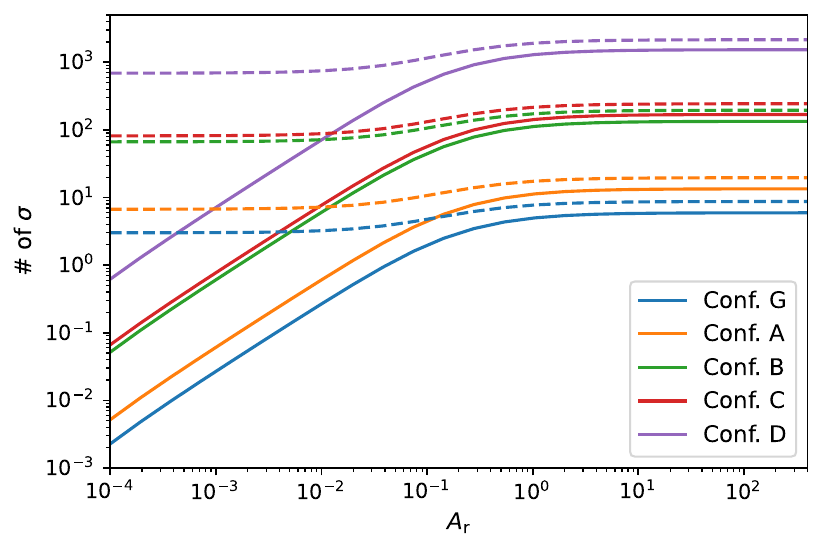}
\caption{The significance (\# of $\sigma$) of the detection (i.e., $1/\delta \beta$) as a function of $A_{\rm r}$, for 21-cm power spectrum measurements. We show two detection scenarios: distinguishing the ERB global signal from the standard case (solid lines), and detecting it relative to zero signal (dashed lines).}
\label{fig:dbeta_power}
\end{figure}

Next we calculate the significance with which the power spectrum of ERB models can be distinguished from the standard cosmological model, or detected (distinguished from zero signal). Fig.~\ref{fig:dbeta_power} shows the significance of the detection (for these two scenarios) as a function of $A_{\rm r}$, for the various observational configurations. Here the significance in both scenarios monotonically increases with $A_{\rm r}$, and it is always easier to detect an ERB signal than to distinguish it from the standard case. For detecting the signal, the significance increases smoothly from the value for the standard case (at low $A_{\rm r}$) to that for the saturated ERB case (at high $A_{\rm r}$), with a transition occurring over the range of $A_{\rm r} \sim 0.1 - 1$. For distinguishing the signal from the standard case, similarly to the global signal, the significance increases roughly as $A_{\rm r}$ until it saturates beyond $\sim A_{\rm r90}$. 

The significance of detecting the ERB power spectrum in these two ways is also listed in Table~\ref{tab:ERBps}, for various values of $A_{\rm r}$ and the various configurations. With the minimal G configuration, the saturated ERB signal can be detected at 8.73$\sigma$ significance, and distinguished from the standard case at 5.94$\sigma$. These values are comparable to the 1,000\,hour global case (Table~\ref{tab:ERB}), and again significantly stronger statistically than the detection of the standard case itself (3.01$\sigma$ in the same configuration). In the A configuration, the significance of the two types is nearly as good as the 10,000\,hour global case, while the B and C configurations exceed the 100,000\,hour global case (and D is much better still). Note that configuration C somewhat outperforms B, due to its higher angular resolution. These comparisons between the statistical strengths of the global signal and power spectrum for the ERB model are generally similar to the comparison for the standard cosmological model (Table~\ref{tab:standard}). 

\begin{table}
\begin{center}
\caption{The significance (\# of $\sigma$) of detecting the ERB 21-cm power spectrum relative to zero signal or the standard signal.}
\label{tab:ERBps}
\begin{minipage}{0.48\textwidth}

\begin{tabular*}
{\textwidth}{@{\extracolsep{\fill}}lcccccc@{\extracolsep{\fill}}}
\hline

 & & \multicolumn{5}{@{}c@{}}{Configuration} \\
\hline
Relative to & $A_{\rm r}$ & G & A & B & C & D \\

\hline
 & 0.001 & 3.03 & 6.76 & 67.1 & 82.2 & 695 \\
 & 0.01 & 3.23 & 7.20 & 71.6 & 87.8 & 747 \\
Zero & 0.1 & 4.76 & 10.7 & 106 & 132 & 1153 \\
 & 0.4 & 6.73 & 15.1 & 150 & 188 & 1661 \\
 & 375 & 8.73 & 19.6 & 195 & 244 & 2156 \\
\hline
 & 0.001 & 0.0267 & 0.0606 & 0.605 & 0.769 & 7.10 \\
 & 0.01 & 0.266 & 0.604 & 6.02 & 7.67 & 70.9 \\
Standard & 0.1 & 1.98 & 4.50 & 44.8 & 57.3 & 530 \\
 & 0.4 & 3.98 & 9.01 & 89.8 & 114 & 1047 \\
 & 375 & 5.94 & 13.4 & 134 & 169 & 1532 \\
\hline
\end{tabular*}
\end{minipage}
\end{center}
\end{table}

For distinguishing the ERB model from the standard case, Table~\ref{tab:sig_para_power} lists the minimum $A_{\rm r}$ values for various levels of significance, for the various observational configurations. For example, configuration G can detect a minimum value of $A_{\rm r} = 1.06$ at 5$\sigma$, and the detection threshold can be improved by an order of magnitude in each step of going to A, to B or C, and then to D, which can go down to $A_{\rm r} = 0.000708$. Generally, the global signal is relatively more sensitive to low values of $A_{\rm r}$ (relative to high $A_{\rm r}$) compared to the power spectrum. This is likely due to the different redshift dependence. Since the thermal noise for the power spectrum measurement goes as the square of system temperature $T_{\rm sys}^2$, it increases much faster with redshift than the signal, unlike the global signal case. Therefore, the power spectrum mostly measures lower $z$, while the global signal can take advantage of higher redshifts, where the ERB signal depends most strongly on the amplitude $A_{\rm r}$. Of course, the global and fluctuation measurements are observationally independent, so ideal would be to have both measurements provide a useful cross-check.

\begin{table}
\begin{center}
\caption{The minimum value of $A_{\rm r}$ (in units of $10^{-2}$) that allows the ERB 21-cm power spectrum to be distinguished from the standard ($A_{\rm r}=0$) case, at various levels of statistical significance.}
\label{tab:sig_para_power}

\begin{minipage}{0.48\textwidth}

\begin{tabular*}
{\textwidth}{@{\extracolsep{\fill}}lccccc@{\extracolsep{\fill}}} 
\hline

& & \multicolumn{4}{@{}c@{}}{Detection limit} \\
\hline
{} & Configuration & 1$\sigma$ & 2$\sigma$ & 3$\sigma$ & 5$\sigma$ \\

\hline
 & G & 4.19 & 10.5 & 21.3 & 106 \\
 & A & 1.69 & 3.60 & 5.92 & 12.1 \\
$A_{\rm r}\,[10^{-2}]$ & B & 0.164 & 0.328 & 0.493 & 0.829 \\
 & C & 0.130 & 0.258 & 0.387 & 0.648 \\
 & D & 0.0153 & 0.0292 & 0.0431 & 0.0708 \\

\hline
\end{tabular*}
\end{minipage}
\end{center}
\end{table}

\section{The millicharged dark matter model}

As noted in the introduction, one way of explaining the tentative EDGES absorption feature is by reducing the baryon temperature through baryon-DM scattering \citep{Barkana:2018}. Any particle physics model that supplies such a new scattering interaction faces additional constraints, such as baryon self-interaction, i.e., a fifth force \citep{Berlin, Barkana2018a}. A model that satisfies these constraints is millicharged dark matter (mDM), in which a small fraction of the dark matter particles have a tiny electric charge, and Coulomb scattering is responsible for the energy transfer \citep{munoz18}. A strong correlation between baryon temperature and the baryon-DM relative streaming velocity tends to imprint large acoustic oscillations on the 21-cm signal \citep{Barkana:2018}, but this signature is erased by drag at early times throughout the mDM parameter space that remains consistent with observational constraints, particularly from the CMB \citep{Kovetz}. It is possible to restore this signature in an interacting millicharged dark matter model \citep{Liu19,imDM}, which is more elaborate (adding a long-range interaction between the millicharged part and the rest of the DM) but also viable over a much wider range of parameters. 

Here we consider the dark ages global 21-cm signal from the simpler, non-interacting mDM model. The parameters of this model are the fraction of the DM mass density that is millicharged ($f_{\rm X}$), the electric charge of the millicharged particles ($\epsilon$, a fraction of the electron charge $e$), and the mass of the millicharged particles ($m_{\rm X}$). We consider five different models with parameter values (see Table~\ref{tab:model_para}) that roughly span the range that is allowed by current constraints and that can explain the EDGES result \citep{Kovetz}. 

\begin{table}
\begin{center}
\caption{The mDM models that we use to illustrate our results, in terms of the model parameters $f_{\rm X}$ (millicharged fraction of the DM), $\epsilon$ (millicharged electric charge) and $m_{\rm X}$ (millicharged particle mass).}
\label{tab:model_para}

\begin{minipage}{0.48\textwidth}

\begin{tabular*}
{\textwidth}{@{\extracolsep{\fill}}lcccccc@{\extracolsep{\fill}}} 
\hline

& \multicolumn{5}{@{}c@{}}{Model} \\
\hline
{} & A & B & C & D & E \\

\hline
$f_{\rm X}$ & 0.004 & 0.004 & 0.004 & 0.001 & 0.001 \\
$\epsilon$ [$10^{-4}$\,e]& 1.0 & 0.1 & 0.1 & 0.3 & 0.1  \\
$m_{\rm X}$ [MeV] & 10 & 3 & 1 & 5 & 1  \\

\hline
\end{tabular*}
\end{minipage}
\end{center}
\end{table}

\begin{figure}
\centering
\includegraphics[width=.48\textwidth, angle=0]{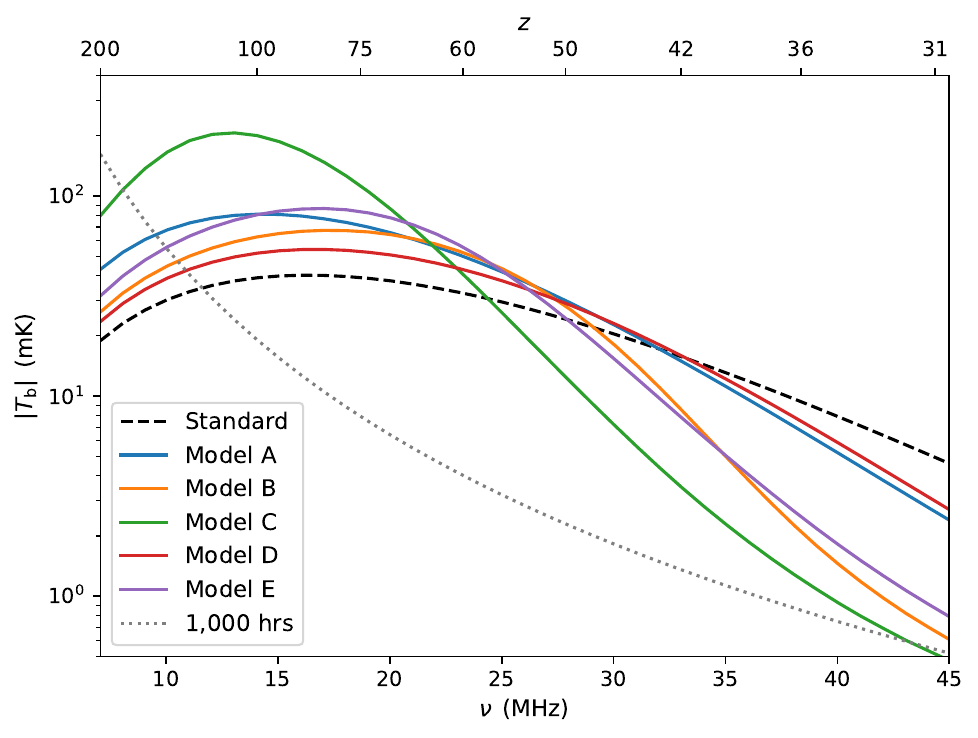}
\caption{The size of the global 21-cm signal as a function of $\nu$ for the standard $\Lambda$CDM model and the mDM models (Table~\ref{tab:model_para}) considered in this work. We also show the expected thermal noise for a global signal experiment observing for integration time 1,000 hrs (grey dotted line) for a bin around each $\nu$ of width $\Delta(\ln \nu) = 1$.}
\label{fig:milli_DM}
\end{figure}

Fig.~\ref{fig:milli_DM} shows the size of the global 21-cm signal from the dark ages as a function of $\nu$ (and $z$), for the mDM models considered in this this work. We also show the standard case for comparison, and the instrumental noise for integration time $t_{\rm int} =$1,000\,hrs for a bin around each $\nu$ of width $\Delta(\ln \nu) = 1$. Unlike the ERB model, the mDM models have a shape versus frequency that is generally similar to that of the standard model, except that the variation with redshift is stronger. In the mDM model, the colder gas has two 21-cm effects: on the one hand, the colder gas (relative to the CMB) tends to produce stronger 21-cm absorption, but on the other hand, colder gas has weaker collisional coupling. At the high-redshift end, where the collisional coupling is above unity (Fig.~\ref{fig:xc}), even lowering it by a small factor has a limited effect, so that the 21-cm global absorption is stronger in the mDM cases than in the standard case. However, at the low-redshift end of the dark ages, the effect on the coupling becomes dominant, and the absorption is actually weaker in mDM than in the standard model (this is reversed later on at cosmic dawn, when the coupling comes from Lyman-$\alpha$ photons). 

As for the ERB model, we calculate two types of statistical significance, for how well each mDM model can be detected (i.e., distinguished from a zero signal), and how well it can be distinguished from the standard signal (i.e., showing that the signal is anomalous and must correspond to exotic physics). As before, we assume observations spanning the redshift range $30-200$ with bins of $\Delta \nu = 1$. Table~\ref{tab:mDM_sig} lists the results. While the global signal in the mDM models varies faster with frequency than the standard signal, this also brings the mDM models closer at high redshifts to the slope of the thermal noise. As a result, the order in terms of which type of detection is easier, varies among the models. Overall, with $t_{\rm int}= 1$,000\,hrs the mDM models can be detected at $4.68-7.16\sigma$ (all higher than the 4.12$\sigma$ for detecting the standard model), and can be distinguished from the standard model at $2.22-9.26\sigma$. 

\begin{table}
\begin{center}
\caption{The significance (\# of $\sigma$) of detecting the global signal in the mDM models, relative to zero signal or the standard signal.}
\label{tab:mDM_sig}
\begin{minipage}{0.48\textwidth}

\begin{tabular*}
{\textwidth}{@{\extracolsep{\fill}}lcccc@{\extracolsep{\fill}}}
\hline

 & & \multicolumn{3}{@{}c@{}}{Integration time} \\
\hline
 & Model & 1,000\,hrs & 10,000\,hrs & 100,000\,hrs \\

\hline
 & A & 4.90 & 15.5 & 49.0 \\
 & B & 5.79 & 18.3 & 57.9 \\
Relative to zero & C & 7.16 & 22.6 & 71.6 \\
 & D & 4.68 & 14.8 & 46.8 \\
 & E & 5.98 & 18.9 & 59.8 \\
\hline
 & A & 3.55 & 11.2 & 35.5 \\
 & B & 5.45 & 17.2 & 54.5 \\
Relative to standard & C & 9.26 & 29.3 & 92.6 \\
 & D & 2.22 & 7.01 & 22.2 \\
 & E & 6.07 & 19.2 & 60.7 \\
\hline
\end{tabular*}
\end{minipage}
\end{center}
\end{table}

\section{Discussion and Conclusions}
\label{sec:conc}
The redshifted 21-cm signal from the dark ages is a powerful cosmological probe with the potential to constrain cosmology. This has been previously shown under the assumption of the standard cosmology. However, there are various studies on non-standard possibilities during the dark ages, which suggest that exotic models could be easier to detect than the standard case. In this paper, we have studied two non-standard models that are consistent with current constraints and that correspond to exotic dark matter properties that go beyond the cold dark matter model: the excess radio background (ERB) model and the millicharged dark matter (mDM) model. We have investigated the effects of these two non-standard models on the redshifted 21-cm signal from dark ages. Both of these models have been mainly motivated by the tentative EDGES detection, but more generally they provide a useful range of possibilities for anticipating potential discoveries once the window of the dark ages is opened up observationally. 

First, we quantified the effects of an ERB on the redshifted 21-cm global signal from the dark ages. We found that the ERB can substantially increase the amplitude of the global signal, depending on the parameter $A_{\rm r}$ (defined as the $z=0$ intensity of the ERB relative to the CMB at 78~MHz). We found that the signal becomes saturated (independent of $A_{\rm r}$) at high $A_{\rm r}$, with 90\% saturation (during the dark ages) reached at $A_{\rm r}=0.4$; this corresponds to 21\% of the minimum value ($A_{\rm r} = 1.9$) required to explain EDGES, and 0.11\% of the value that would explain the entire observed extragalactic radio background.

Using Fisher analysis, we forecast the detection significance of the ERB signal. For 1,000\,hrs of integration of the global signal, the 90\% saturation case can be detected at 5.89$\sigma$ significance, and it can be distinguished from the standard signal at 8.45$\sigma$; these are substantially stronger results than the detection of the standard signal itself (4.12$\sigma$ in this case), due to the greater amplitude of the ERB signal, and accounting (optimistically) for degeneracy with the foreground. A much smaller value of $A_{\rm r}$ can be distinguished from the standard signal at 5$\sigma$: $A_{\rm r} = 0.0389$, which is only 2.0\% of the minimum value for explaining EDGES, and just 0.010\% of the extragalactic radio background. All these results get much stronger if the integration time is increased significantly beyond 1,000\,hrs, so that the number of $\sigma$ is an order of magnitude larger for $t_{\rm int}= 100$,000\,hrs, which is feasible to achieve with an array of global antennas; with this larger integration time, the amplitude that can be distinguished from the standard signal at 5$\sigma$ is $A_{\rm r} = 0.00175$, which is $9.2 \times 10^{-4}$ of the minimum value for EDGES, and 
$4.7 \times 10^{-6}$ of the extragalactic radio background.

We also studied the 21-cm power spectrum in the ERB model. As in the case of the standard model, compared to the global signal it would take a much larger effort to reach significant results for the ERB model with 21-cm fluctuations from the dark ages, but in the long-run, much better results are possible. Similarly to the global signal, the 21-cm power spectrum rises with $A_{\rm r}$ and becomes saturated at $A_{\rm r} \gtrsim 0.4$. Here, though, the thermal noise rises with redshift as the square of system temperature, which is significantly faster than the signal, while the rates are similar for the global signal, allowing the latter to benefit more from the highest redshifts. For our minimal G configuration of a dark ages interferometric array, the 90\% saturation case can be detected at 6.73$\sigma$ significance, and it can be distinguished from the standard signal at 3.98$\sigma$; these are again substantially stronger results than the detection of the standard signal itself (3.01$\sigma$ in this case). The value of $A_{\rm r}$ that can be distinguished from the standard signal at 5$\sigma$ is $A_{\rm r} = 1.06$, as the power spectrum is less sensitive to low values of $A_{\rm r}$ compared to the global signal (due to the different redshift behavior). This constraint for the B and C configurations would be comparable to that for a 10,000\,hr global experiment, and the D configuration would go down to $A_{\rm r} = 0.000708$, lower by a factor of 2.47 than the value achievable with a 100,000\,hr global experiment. The power spectrum is much less effective on this measure compared to the significance of detection (of either the standard signal or the 90\% saturated ERB model relative to zero signal or the standard signal); for the latter, the B and C configurations perform better than 100,000\,hr global, and D gives a further order of magnitude improvement in the number of $\sigma$. Of course, it would be best to pursue both global and power spectrum measurements, as they would be observationally independent and thus provides complementary information and a powerful cross-check.

Finally, we investigated the global signal of the redshifted 21-cm line from the dark ages in the model of gas-DM cooling with mDM. We considered five different models with parameter values that span the region allowed by current constraints and that can also explain the EDGES anomaly. We found that the shape and amplitude of the global signal are significantly different in these models compared to the standard case. As a result, we showed that the detection of the mDM signal is feasible with future dark ages global signal observations, with greater significance than the detection of the standard signal. Also, the mDM models can be distinguished from the standard signal with comparable significance in most cases. For example, the mDM Model C, which has the strongest signal among the models considered, can be distinguished from zero signal with a significance of $7.16\sigma$ and from the standard signal with a significance of $9.26 \sigma$ for 1,000\,hrs of integration time.

Overall, we have shown that the exotic dark matter models can have a significant impact on the 21-cm global signal and power spectrum from the dark ages. Our results suggest that future observations of the 21-cm signal will be able to detect or constrain these exotic models, and thus provide valuable insights on fundamental cosmology.

\section*{Acknowledgements}
We greatly acknowledge discussions with David Neufeld and Jens Chluba regarding CMB heating. RM is supported by the Israel Academy of Sciences and Humanities \& Council for Higher Education Excellence Fellowship Program for International Postdoctoral Researchers. RM and RB acknowledge the support of the Israel Science Foundation (grant No. 2359/20). AF is supported by Royal Society University Research Fellowship \#180523.

\section*{Data availability}
The data underlying this article are available upon request from the corresponding author.

\bibliographystyle{mnras} 
\bibliography{refs}


\vfill
\bsp
\label{lastpage}
\end{document}